\documentclass[preprint,aps,showkeys]{revtex4}
\usepackage{epsfig}
\begin{document}
\title{On the nonextensivity of the long range X-Y model}
\author{Ra\'ul Toral}
\email{raul@imedea.uib.es} 
\affiliation{Instituto Mediterr\'aneo
de Estudios Avanzados (IMEDEA), CSIC-Universitat de les Illes
Balears\\ Ed. Mateu Orfila, Campus UIB, 07122 Palma de Mallorca, Spain}
\homepage{http://www.imedea.uib.es}

\date{\today}

\begin{abstract} 
It will be given analytical and numerical evidence supporting that the X-Y model yields an extensive, i.e. proportional to the number of degrees of freedom $N$, internal energy $U$ for any value of the interaction range.
\end{abstract}

\maketitle
The X-Y is a well known model of Statistical Mechanics. It is defined by a set of conjugate pairs of variables $(\theta_i,p_i)_{i=1,\dots,N}$, $N$ being the number of degrees of freedom, with a Hamiltonian function
\begin{equation}
\label{ham}
{\cal H} =\frac{1}{2}\sum_{i=1}^Np_i^2+\frac{\kappa}{2}\sum_{i,j=1}^N\frac{1-\cos(\theta_i-\theta_j)}{r_{ij}^{\alpha}}\equiv K+V,
\end{equation}
where the kinetic, $K$, and potential, $V$, contributions to the Hamiltonian have been identified. In this expression $r_{ij}$ is the distance between the sites $(i,j)$ of a regular lattice of dimensionality $d$ and lattice spacing equal to one. For example, in the linear case that will be analysed in detail later, $d=1$, it is $r_{ij}=|i-j|$. Periodic boundary conditions are assumed, and it is understood the convention of minimal distance between sites $i$ and $j$. The constant $\kappa$ sets the interaction strength. The definition is such that, for ferromagnetic coupling $\kappa\ge 0$ (the only one considered in this paper) the minimum value of the energy is equal to zero.

The parameter $\alpha$ sets the interaction range and its variation allows to recover some limits of interest. For example, the case $\alpha=\infty$, for which $r_{ij}^{-\alpha}=0$ unless $r_{ij}=1$, is the nearest neighbor version, whereas the opposite limit, $\alpha=0$, is a mean field Hamiltonian in which all sites interact equally. This mean field limit is usually studied\cite{ST71} under the condition that the interaction strength is of order $N^{-1}$, i.e. by setting $\kappa=\epsilon/N$ with $\epsilon=O(1)$. It is known that in this case the model exhibits a phase transition from an ordered to a disordered phase at a critical value $\beta_c=2/\epsilon$. This result follows\cite{AR95} from an analysis of the partition function 
\begin{equation}
{\cal Z}(\beta,N)=\int_{-\infty}^{\infty}dp_1\dots\int_{-\infty}^{\infty}dp_N\int_{0}^{2\pi}d\theta_1\dots\int_{0}^{2\pi}d\theta_N\,{\rm e}^{-\beta{\cal H}}
\end{equation}
at inverse temperature $\beta=1/k_BT$, as usual.

The model has received recently some attention\cite{TA00} for a value of the interaction strength $\kappa=O(1)$ as an example of a non--extensive system. It has been argued\cite{TS95} that in this case the internal energy $U=\langle {\cal H}\rangle$, as well as other thermodynamical potentials, should scale as $U\sim N^{2-\alpha/d}$ for $\alpha < d$ (the so--called {\sl nonextensive} regime), whereas for $\alpha > d$, the {\sl extensive} regime $U\sim N$ is recovered. 

In this paper, I will present arguments showing that the internal energy is an extensive function $U(N,T)=Nu(T)$ for any value of $\alpha$. The result is based upon an exact calculation of the partition function in the mean--field limit $\alpha=0$ as well as some numerical simulations. 

Let me first sketch the proof for $\alpha=0$. Introducing the total ``spin vector"  \makebox{$\vec M = \sum_i \vec m_i$} with \makebox{$\vec m_i = (\cos \theta_i,\sin\theta_i)$} it is possible to write the potential energy as \makebox{$V=\frac{\kappa}{2}(N^2-M^2)$}. Using this relation and the Hubbard-Stratonovich transformation, one arrives at the following expression for the partition function\cite{arfor}:
\begin{equation}
{\cal Z}=\left(\frac{2\pi}{\beta}\right)^{\frac{N}{2}}{\rm e}^{-\beta\frac{\kappa}{2}N^2}\frac{(2\pi)^N}{\beta\kappa}\int_0^{\infty}dy\,y {\rm e}^{-\frac{y^2}{2\beta\kappa}}\left[I_0(y)\right]^N\equiv {\cal Z}_K{\cal Z}_V
\end{equation}
where ${\cal Z}_K=\left(\frac{2\pi}{\beta}\right)^{\frac{N}{2}}$ and ${\cal Z}_V$ are, respectively, the kinetic and potential contributions and $I_n(y)$ is the modified Bessel function of order $n$. The standard calculation\cite{AR95} assumes that $\kappa = O(N)$ and finds the dominant saddle point contribution to the above integral in the limit $N\to \infty$. A similar calculation can be carried out assuming that $\kappa=O(1)$. In this case, the location of the saddle point $y_0$ is given as the solution of
\begin{equation}
\frac{1}{Ny_0}-\frac{2y_0}{\beta\kappa N}+\frac{I_1(y_0)}{I_0(y_0)}=0.
\end{equation}
This can be solved in the limit $N\to\infty$, with the result that \makebox{$y_0=\frac{N\beta\kappa}{2}-\frac{1}{2}+O(N^{-1})$}. The resulting partition function is:
\begin{equation}
{\cal Z}=\left(\frac{2\pi}{\beta}\right)^{\frac{N}{2}}\frac{1}{\beta\kappa}
\left(\frac{2\pi}{\beta\kappa N}\right)^{\frac{N}{2}}
\end{equation}
It turns out that the Helmholtz free energy $F=-k_BT\ln{\cal Z}$ is not an extensive function since it contains terms of order $N\ln N$. However, these terms disappear when computing the internal energy $U=-\frac{\partial \ln {\cal Z}}{\partial \beta}$. This yields \makebox{$U=(N+1)/\beta$}. In the thermodynamic limit the dominant term is:
\begin{equation}
\label{e0}
U=\frac{N}{\beta}
\end{equation}
showing that the internal energy is indeed an extensive function of the number of degrees of freedom. Note that the kinetic and potential terms in the Hamiltonian contribute equally, $\displaystyle \langle K\rangle=\langle V\rangle = \frac{N}{2\beta}$, to the internal energy.

The nearest neighbor interaction, equivalent to the limit $\alpha=\infty$, has been studied in Ref.\cite{LPR87,EKL94} for the linear lattice $d=1$. It is found that the internal energy is again an extensive function, namely:
\begin{equation}
\label{e1}
U=N\left[\frac{1}{2\beta}+\kappa\left(1-\frac{I_1(\beta\kappa)}{I_0(\beta\kappa)}\right)
\right]
\end{equation}
 Therefore, since the limits $\alpha=0$ and $\alpha=\infty$ both yield an extensive internal energy, it is natural to speculate that a similar behavior will hold for any value of the parameter $\alpha$. In order to check this result I have performed Monte Carlo simulations of the one dimensional system given in (\ref{ham}) for the values of $\alpha=0,0.5,\infty$ and a value of the coupling strength $\kappa=1$. Whereas the limiting cases $\alpha=0,\infty$ are only for the purposes of checking that the numerical simulations do reproduce the theoretical results, the intermediate value $\alpha=0.5$ can not be compared with any a priori theoretical calculation. 
  
I have used a standard Metropolis algorithm\cite{BH88} to sample the phase space with a probability proportional to ${\rm e}^{-\beta {\cal H}}$: (i) Choose randomly a lattice site $i$. (ii) Propose a change $(p_i,\theta_i)\to (p_i',\theta_i')$ of the variables associated to this site; the proposed values for $p_i'$ and $\theta_i'$ are selected randomly from a uniform distribution in $(p_i-\delta_p, p_i+\delta_p)$ and $(\theta_i-\delta_{\theta}, \theta_i+\delta_{\theta})$, respectively. (iii) Accept the change with a probability $\min[1,{\rm e}^{-\beta\Delta {\cal H}}]$, being $ \Delta {\cal H}={\cal H'-H}$ the change in energy involved in the proposed change. The values of $\delta_p$ and $\delta_{\theta}$ are chosen such that acceptance probability remains close to 50\%. Typically, after $10^5$ Monte Carlo steps per particle for equilibration, results have been averaged for $10^6$ Monte Carlo steps per particle for each value of the inverse temperature $\beta$. In all cases, the following system sizes have been taken: $N=50,100,200$.

\begin{figure}[h]
\centerline{\epsfig{figure=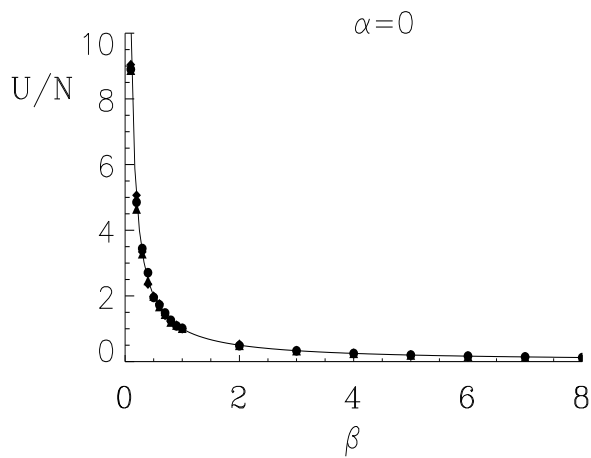,width=9.5cm,height=7cm,angle=0}}
\caption{\label{fig1}Internal energy $U$ rescaled by the number of degrees of freedom for $\alpha=0$ and coupling strength $\kappa=1$. The symbols are the result of the Monte Carlo simulations, whereas the solid line is the theoretical prediction Eq. (\ref{e0}). Circles, diamonds and triangles correspond, respectively to $N=50,100,200$. It is observed a very good collapse of the data for all values of $N$ showing the extensive character of the internal energy.}
\end{figure}

\begin{figure}[h]
\centerline{\epsfig{figure=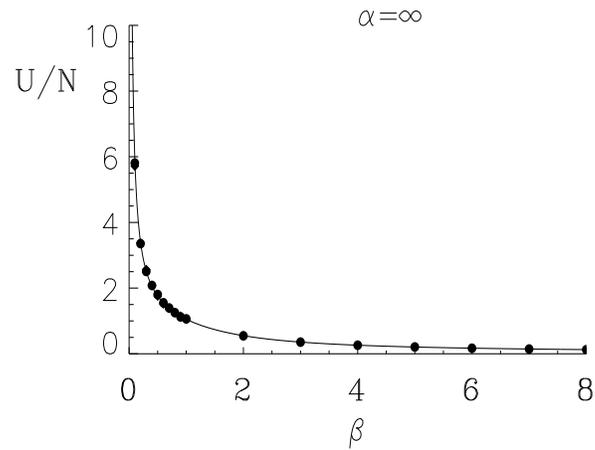,width=9.5cm,height=7cm,angle=0}}
\caption{\label{fig2}Similar as Fig. \ref{fig1} for $\alpha=\infty$. The symbols are the results of the numerical simulations (same symbol meaning than in Fig.\ref{fig1}) and the solid line is the theoretical prediction Eq.(\ref{e1}).}
\end{figure}

\begin{figure}[h]
\centerline{\epsfig{figure=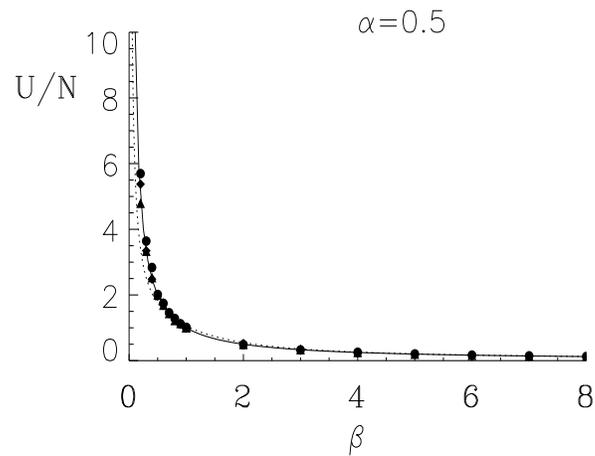,width=9.5cm,height=7cm,angle=0}}
\caption{\label{fig3}Similar as Fig. \ref{fig1} for $\alpha=0.5$. The symbols are the results of the numerical simulations (same symbol meaning than in Fig.\ref{fig1}). For comparison, the theoretical predictions for $\alpha=0$, Eq. (\ref{e0}), solid line, and $\alpha=\infty$, Eq.(\ref{e1}), dotted line, are also included.}
\end{figure}

\begin{figure}[h]
\centerline{\epsfig{figure=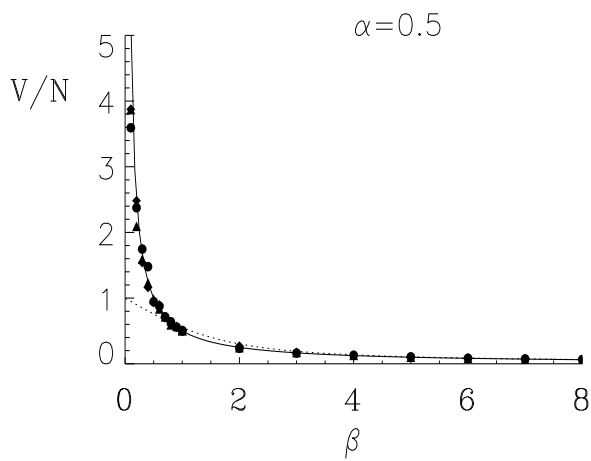,width=9.5cm,height=7cm,angle=0}}
\caption{\label{fig4}The potential part $V$ of the internal energy in the same case than in figure \ref{fig3}.  The symbols are the results of the numerical simulations (same symbol meaning than in Fig.\ref{fig1}) whereas the solid and dotted lines are the theoretical predictions for the limit cases $\alpha=0$ and $\alpha=\infty$, respectively.}
\end{figure}

Figures \ref{fig1}-\ref{fig3} plot the internal energy rescaled by the number of degrees of freedom, $U/N$, as a function of the inverse temperature $\beta$, for the differer values of $\alpha$ considered.. Figure \ref{fig4} plots the rescaled potential energy, $V/N$, in the case $\alpha=0.5$. In all cases the data for all values of $N$ collapse in a single curve, showing the extensive nature of the internal energy for all the values of $\alpha$. Moreover, the cases $\alpha=0$ and $\alpha=\infty$, Figs. \ref{fig1} and \ref{fig2}, follow the theoretical predictions given by Eqs. (\ref{e0}) and (\ref{e1}), respectively. Note that, as shown in Figs. \ref{fig3}, there is not much quantitative difference between the values of the internal energies corresponding to these two limiting cases.
  
\begin{figure}[h]
\centerline{\epsfig{figure=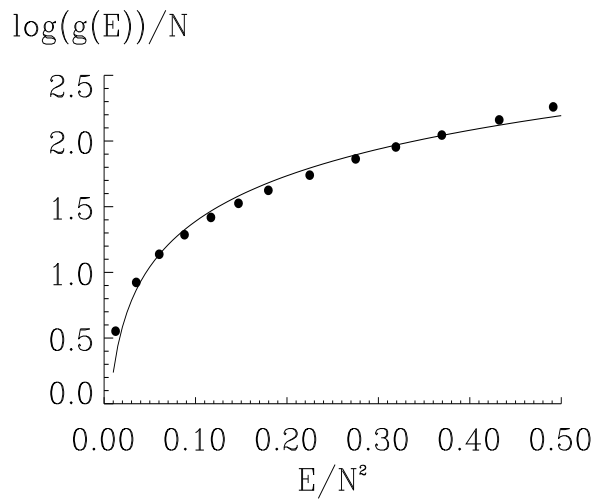,width=9.5cm,height=7cm,angle=0}}
\caption{\label{fig5}Density of states g(E) for the potential part of the X-Y Hamiltonian in the case $\alpha=0$. The solid line is the theoretical prediction Eq. (\ref{e8}), with a constant $c=2.54$, while the dots are the result of a numerical calculation \cite{STP02} of the same quantity. Please note that in reference \cite{STP02} the data were plotted using base 10 logarithms instead of the natural logarithms used here.}
\end{figure}

As other authors have found\cite{TA00}, it is also possible to map the results of the X-Y model with $\kappa=O(1)$ to the same model with $\kappa=O(N)$ if one allows to rescale the temperature $T$ by the number of degrees of freedom $N$. In this way it is possible to achieve a sort of non--extensive regime in which $U(N,T)=N^2u(T/N)$ (for $\alpha=0$). The results of this paper, however, concern the true thermodynamic limit, $N\to\infty$, while keeping $T=O(1)$. It has been shown that this limit yields an extensive internal energy and that this extensive behavior can be recovered by simple Monte Carlo simulations. It is interesting to remark that the same extensive behavior will be found in microcanonical simulations. In fact, it is possible to compute the density of states $g(E)$ for the potential part of the Hamiltonian as the inverse Laplace transform of the partition function ${\cal Z}_V$\cite{PA96}. This gives:
\begin{equation}
\label{e8}
\frac{\log(g(E))}{N}=\frac{1}{2}\log\left(\frac{E}{N^2}\right)+ c
\end{equation}
where $c$ is a constant of order one. As shown in figure \ref{fig5}, this result is fully consistent with a numerical calculation of the density of states\cite{STP02} which predicted empirically a form $g(E)={\rm e}^{N\phi(E/N^2)}$. Within this microcanonical ensemble, the internal (potential) energy is found by using $\beta=\frac{\partial \ln g(E)}{\partial E}$ or $E=N/2\beta$, the same result we obtain before for the average potential energy $\langle V\rangle$.

The existence of this extensive regime can be considered surprising at first sight, since the interaction terms are very strong. From the mathematical point of view, the extensive regime is due to the continuous nature of the excitations of the ground state of the Hamiltonian (\ref{ham}). A similar analysis for the long range Ising model in  which the spins can only vary by an integer amount indeed produces a non-extensive internal energy. In the X-Y model with long range interactions and $\kappa=O(1)$ there is no phase transition at any finite value of the temperature in the thermodynamic limit, and the system is always in an ordered situation. In fact, the order parameter is $M=O(N)$ for all values of the temperature.  

In summary, it has been given analytical and numerical evidence showing that the internal energy derived from the X-Y Hamiltonian is an extensive function of the number of degrees of freedom, $U(N,T)=Nu(T)$, for any value of the parameter $\alpha$ setting the interaction range. Other thermodynamic potentials, such as the Helmholtz free energy, contain terms of the order $N\log N$. Non extensive regimes require to scale the temperature with $N$.

{\bf Acknowledgments}

This work is supported by the Ministerio de Ciencia y Tecnolog{\'\i}a (Spain)
and FEDER, projects BFM2001-0341-C02-01 and BFM2000-1108.

\begin{thebibliography}{99}
\bibitem{arfor} This is nothing but formula (14) of \cite{AR95} setting $\epsilon=N\kappa$.
\end{thebibliography}
\end{document}